\documentclass{aa}
\usepackage{hyperref}

\usepackage{graphicx}
\usepackage{txfonts}
\usepackage{gensymb}
\usepackage{ulem}
\usepackage{siunitx}
\usepackage{natbib}
\bibpunct{(}{)}{;}{a}{}{,} 
\hypersetup{%
  colorlinks=true,
  linkcolor=blue,
  urlcolor=blue,
  citecolor=blue
  }

\makeatletter
\renewcommand*\aa@pageof{, page \thepage{} of \pageref*{LastPage}}
\makeatother

\newcommand{\hrieuv}{HRI\textsubscript{EUV}\xspace}
\newcommand{\hrilya}{HRI\textsubscript{Ly$\alpha$}\xspace}
\newcommand{\fsi}{FSI\xspace}

\begin{document}

    \title{Temperature of quiet Sun small scale brightenings observed by EUI on board Solar Orbiter: Evidence for a cooler component}
   \author{%
        {A. Dolliou} \inst{\ref{aff:ias}}
        \and {S. Parenti} \inst{\ref{aff:ias}}
        \and {F. Auchère} \inst{\ref{aff:ias}}
        \and {K. Bocchialini} \inst{\ref{aff:ias}}
        \and {G. Pelouze} \inst{\ref{aff:ias}}
        \and {P. Antolin} \inst{\ref{aff:northumbria}}
        \and {D. Berghmans} \inst{\ref{aff:rob}}
        \and {L. Harra} \inst{\ref{aff:davos},\ref{aff:eth}}
        \and {D. M. Long} \inst{\ref{aff:ucl}}
        \and {U. Schühle} \inst{\ref{aff:mpss}}
        \and {E. Kraaikamp} \inst{\ref{aff:rob}}
        \and {K. Stegen} \inst{\ref{aff:rob}}
        \and {C. Verbeeck} \inst{\ref{aff:rob}}
        \and {S. Gissot} \inst{\ref{aff:rob}}
        \and {R. Aznar Cuadrado} \inst{\ref{aff:mpss}}
        \and {E. Buchlin} \inst{\ref{aff:ias}}
        \and {M. Mierla} \inst{\ref{aff:rob},\ref{aff:igra}}
        \and {L. Teriaca} \inst{\ref{aff:mpss}}
        \and {A. N. Zhukov} \inst{\ref{aff:rob},\ref{aff:sinp}}
          }

   \institute{%
        \label{aff:ias}{Université Paris-Saclay, CNRS,  Institut d'astrophysique spatiale, 91405, Orsay, France}
        \and \label{aff:northumbria}{Department of Mathematics, Physics and Electrical Engineering, Northumbria University, Newcastle Upon Tyne, NE1 8ST, UK}
        \and \label{aff:rob}{Solar-Terrestrial Centre of Excellence – SIDC, Royal Observatory of Belgium, Ringlaan -3- Av. Circulaire, 1180 Brussels, Belgium}
        \and \label{aff:davos}{Physikalisch-Meteorologisches Observatorium Davos, World Radiation Center, 7260, Davos Dorf, Switzerland} 
        \and \label{aff:eth}{ETH-Zürich, Wolfgang-Pauli-Str. 27, 8093 Zürich, Switzerland}
        \and \label{aff:ucl}{UCL-Mullard Space Science Laboratory, Holmbury St. Mary, Dorking, Surrey, RH5 6NT, UK}
        \and \label{aff:mpss}{Max Planck Institute for Solar System Research, Justus-von-Liebig-Weg 3, 37077 Göttingen, Germany}
        \and \label{aff:igra}{Institute of Geodynamics of the Romanian Academy, Bucharest, Romania}
        \and \label{aff:sinp}{Skobeltsyn Institute of Nuclear Physics, Moscow State University, 119992 Moscow, Russia}
        \\ \email{antoine.dolliou@universite-paris-saclay.fr}
             }

    \date{Received 7 September 2022 / 
    Accepted 4 January 2023}

 
  \abstract
   {On May 30, 2020, small and short-lived extreme-UV (EUV) brightenings in the quiet Sun were observed  over a four-minute sequence by the EUV channel of the Extreme Ultraviolet Imager - High Resolution Imager (EUI\hrieuv) on board the Solar Orbiter. The brightenings' physical origin and possible impact on coronal or transition region (TR) heating are still to be determined.}
   {Our aim is to derive the statistical thermal evolution of these events in order to establish their coronal or TR origin.}
   {Our thermal analysis took advantage of the multithermal sensitivity of the Atmospheric Imaging Assembly (AIA) imager on board the Solar Dynamics Observatory. We first identified the \hrieuv events in the six coronal bands of AIA. We then performed a statistical time lag analysis that quantified the delays between the light curves from different bands, as these time lags can give significant insight into the temperature evolution of the events. The analysis was performed taking into account the possible contribution of the background and foreground emissions to the results.}
   {For all nine couples of AIA bands analyzed, the brightening events are characterized by time lags inferior to the AIA cadence of \SI{12}{\second}. Our interpretation for these short time lags is the possible copresence of events that reach or do not reach coronal temperatures ($\approx$ \SI{1}{\mega\kelvin}). We believe that the cool population dominates the events analyzed in this work.}
   {}

\titlerunning{Temperature of EUI QS small-scale brightenings: Evidence for a cooler component}
\authorrunning{Dolliou et al.}
   
\maketitle
%

\section{Introduction}

Decades of investigation suggest that the solar corona is formed and maintained through small-scale processes, even though the mechanisms at the origin of such processes are only partially understood. 
Wave dissipation and magnetic field reconnection are present in the solar atmosphere and are the main candidate processes for the solar corona's plasma heating. See for instance \cite{Reale2014} and \cite{Viall21} for a review on the argument.

Coronal observations suggest that the dissipation of magnetic energy leading to coronal heating must happen at unresolved spatial scales, and while many dissipation mechanisms are impulsive in nature, it is unclear whether the dissipation has a more continuous or sporadic character on average. The properties of the coronal heating events, such as their amplitude, duration, and frequency, are still a matter of debate.

\cite{Parker1988} proposed magnetic reconnection as the origin of these unresolved heating events (which became known as nanoflares). His theory is based on the shuffling and intermixing of the photospheric footpoints of magnetic flux tubes, which would produce reconnection and subsequent formation of tiny current sheets in which the energy is dissipated.
This idea has been generalized in recent years, particularly for active region heating where processes other than reconnection (wave propagation) may also be at the origin of the nanoflares energy \citep{VanDoorsselaere2020,Viall21}. 
For instance, small-scale energy dissipation can occur through a turbulent cascade created by the interaction of nonlinear waves \citep[e.g.][]{Buchlin&Velli2007} or through shock heating from nonlinear mode conversion \citep[][]{Moriyasu2004}.

Studies addressing the heating of the quiet Sun (QS) indicate that waves and reconnections are also present \citep[e.g.][]{McIntosh2011,Hahn&Savin2014, Upendran2021,Upendran&Tripathi2022}. 
In addition, observations of the corona from the hard X-rays \citep[e.g.][]{Crosby1993,Shimizu1995,Hannah&Hudson2010} to the UV bands \citep[e.g.][]{Berghmans&Clette&Moses1998, harra_2010,Aschwanden&Parnell2002} also suggest that small-scale impulsive heating may play a role here.
These observations have revealed that unresolved small bright transient events increase in number everywhere in the corona any time the spatial and temporal resolutions of instruments are increased. 

Examples of small and fast phenomena in the corona have been observed during the High-Resolution Coronal Imager (Hi-C) sounding rocket flights \citep{Kobayashi2014}, during which images were recorded in a band centered on 193~\AA \ (including the Fe XII 195~\AA\ line). These observations were made with a spatial resolution of about $0.3^{\prime \prime}$ \citep[$\approx 220$ km,][]{Winebarger_2014}. 
The Hi-C instrument resolved small cool loops \citep{Winebarger2013} and extreme-UV (EUV) bright dots with characteristic lengths of 680~km, durations of \SI{25}\second,\ and temperatures ranging between 0.5 and \SI{1.5}{\mega\kelvin} \citep{Regnier2014}. 

The Interface Region Imaging Spectrograph \citep[IRIS; ][]{DePontieu2014} reaches a resolution of $\approx 0.33^{\prime\prime}$ -- $0.4^{\prime\prime}$ ($\approx$ 240 -- \SI{290}{\kilo\meter} in the corona) but is mostly sensitive to the transition region (TR) and chromospheric temperatures.
With IRIS and the Atmospheric Imaging Assembly \citep[AIA; ]{Lemen2012} on board the Solar Dynamics Observatory \citep[SDO; ][]{Pesnell2012}, it was possible, for instance, to observe tiny, short-lived, and multithermal {"nanojets"} \citep[size 1000 -- \SI{2000}{\kilo\meter}, $\sim$\SI{15}{\second}, with chromospheric to coronal temperatures,][]{antolin_reconnection_2021, Sukarmadji_2022ApJ...934..190S} in large cool loops, which were interpreted as the transverse motion of field lines reconnecting at small angles. Larger jet-like structures \citep{Innes&Teriaca2013} were detected  with the Solar Ultraviolet Measurements of Emitted Radiation spectrometer \citep{Wilhelm1995} on board the Solar and Heliospheric Observatory (SOHO), along with UV \citep{Peter2014} and  EUV \citep{Young2018} bursts.
IRIS has also observed "unresolved fine structures" in TR lines, which have been associated with short ($\approx$ 4 --  \SI{12}{\mega\meter}) loops or parts of loops. They were seen at the limb in QS regions and shown to be highly variable (a few minutes in lifetime), with strong Doppler shift dynamics (up to \SI{100}{\kilo\metre\per\second}). 

In addition to the aforementioned sporadic and short duration \mbox{Hi-C} rocket flights, SDO/AIA is able to obtain full-Sun images with a resolution of $1.5^{\prime \prime}$ (corresponding to $\approx$ \SI{1100}{\kilo\meter} in the corona). Using the AIA 171 and 193 \AA\ channels, \cite{2014ApJ...787..118R} detected small jets ("jetlets") at the footpoint of coronal plumes. More recently, \cite{chitta2021} characterized the statistical properties of small EUV bursts detected in AIA 171, 193, and 211~\AA \ sequences. Similar and smaller scales brightening are now observed by Solar Orbiter. 

The Solar Orbiter mission \citep[][]{Muller, Zouganelis2020} carries, as part of the remote-sensing payload \citep{Auchere2020}, the Extreme Ultraviolet Imager (EUI) suite \citep{EUI_instrument}. The High-Resolution Imager (\hrieuv) and the Full Sun Imager (\fsi) 174 channels are dominated by emission from lines of \ion{Fe}{IX} and \ion{Fe}{X}. They image the plasma emission of the high TR and corona, which is the region of interest for this work. At its closest, the Solar Orbiter approaches the Sun down to 0.28~AU, allowing a two-pixel spatial resolution of $\approx 200$ km of the corona, with a maximal cadence of \SI{1.6}{\second}, thus providing the highest spatial and temporal resolution images to date at these wavelengths, for extended periods of time and on a variety of targets.

On May 30, 2020, when the Solar Orbiter was at 0.556~AU, \hrieuv made its first observation of the QS corona at high spatial (\SI{400}{\kilo\metre}) and temporal (\SI{5}{\second} of cadence) resolutions. During this four-minute sequence, 1467 small EUV brightenings of variable size (400 to \SI{4000}{\kilo\meter}) and lifetime (10 to \SI{200}{\second}) were detected, referred to as "campfires" \citep{Berghmans2021}. The \hrieuv field of view was also visible by SDO/AIA, and part of the events detected by \hrieuv were also visible in at least one of the AIA coronal bands because of the lower spatial and temporal resolutions of AIA (about \SI{1100}{\kilo\meter} and \SI{12}{\second}, respectively). \cite{Berghmans2021} used the AIA observations to infer the temperature of the EUV brightenings, applying the differential emission measure diagnostic method of \cite{Hannah&Kontar2012}. The resulting distribution was centered around \SI{1.3}{\mega\kelvin}. 

These features are yet to be better characterized, but initial investigations suggest that their origin is linked to photospheric magnetic cancellation \citep{Panesar_2021} or magnetic reconnection close to the TR or the chromospheric part of the loops \citep{2022A&A...660A.143K}.
\cite{Zhukov2021} found that these EUV brightenings are low lying (\SIrange{1}{5}{\mega\metre}), which indicates that they could be chromospheric or transition region features. 
The authors noticed that the estimated heights of the features are larger than their apparent lengths. If these events are loops, this implies that \hrieuv does not see their full extent. Therefore, if they reach \SI{1}{\mega\kelvin}, they do so only at their apex.

\cite{Winebarger2013}, using Hi-C and SDO/AIA data, estimated the temperature of small inter-moss loops to be about $2.8\times 10^5\,\mathrm{K}$. These had a projected length between about 5 and \SI{7}{\mega\metre,} and their light curves, from the different AIA bands, peaked at the same time, suggesting the absence of cooling from coronal temperature. These Hi-C loops are larger than the ones observed by \cite{Berghmans2021} and \cite{Zhukov2021}. Furthermore, they were observed in active regions. However, it is possible that they share similar physical mechanisms.

These results motivated our work to further investigate the thermal properties of the \hrieuv events. We performed a statistical study of over 1000 detected events, and the rest of the QS was used as a reference (see Sect. \ref{data}). 
Our analysis is based on the time lag method (see Sect. \ref{method}) applied to the AIA light curves from several pairs of channels. This method has been extensively used in active regions to study loops submitted to thermal non-equilibrium \citep[][]{,Froment_2015,2017ApJ...835..272F,Froment2020,froment:tel-01402981} and to test the nanoflare theory \citep{Viall_2011,2012ApJ...753...35v,Viall_2015,viall_survey_2017}.
The novelty of the present work relies on the application of this technique to QS region data and over short 
time lags.
In Sect. \ref{results}, we show that there is no or little sign of a lag between all the chosen AIA bands. The implications of these results are discussed in Sect. \ref{discussion}.

\section{Observations and data reduction}
\label{data}

\begin{figure*}
    \centering
    \includegraphics[width=\textwidth]{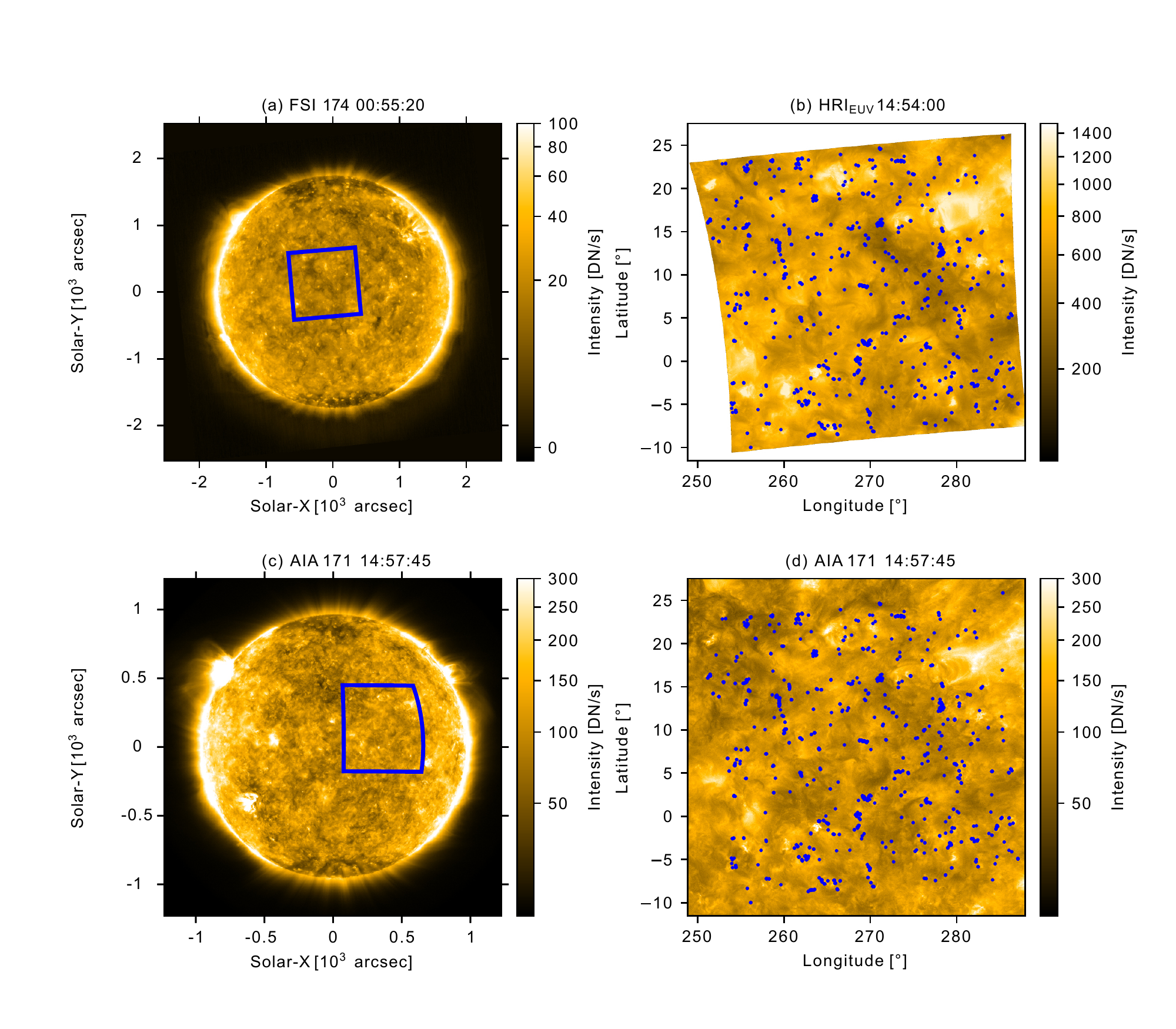}
    \caption{Images captured on May 30, 2020. {Upper row:} Field of view observed by \fsi 174  (a) and the first image of the \hrieuv sequence (b) in Carrington coordinates. The \fsi image is the closest available to the \hrieuv sequence. {Lower row:} AIA 171 image (c) and remapped on the same grid as \hrieuv (d). Blue rectangles in the left column correspond to the field of view on the right column, and the blue dots in the right column are the positions of the 1467 detected events. }
    \label{fig::field_of_view_label_modified}
\end{figure*}
On May 30, 2020, while the Solar Orbiter mission was still performing commissioning activities, \hrieuv observed a QS region at a cadence of 5 seconds for 4 minutes from 14:54:00\,to 14:58:05\,UT. The field of view of \hrieuv  is visible in a full-Sun image taken in the FSI 174 channel (Fig.~\ref{fig::field_of_view_label_modified} (a)). Fig.~\ref{fig::field_of_view_label_modified} (c) shows the corresponding field of view on a full-Sun image of AIA 171, as observed by SDO, which has a similar temperature response, peaking at \SI{0.9}{\mega\kelvin}.
The  apparent difference in position of the \hrieuv field of view between Fig.~\ref{fig::field_of_view_label_modified} (a) and (c) was caused by the separation angle, equal to $31.5^{\degree}$, between the Solar Orbiter line of sight and the Sun-Earth line. 

\subsection{Detection of the EUV brightenings by HRIEUV}
\label{sec::detectionHRI}

The \hrieuv data used for the present work\footnote{EUI Data Release 1.0 \url{https://doi.org/10.24414/wvj6-nm32}} was taken at 0.556~AU from the Sun, resulting in a spatial resolution of $\sim\SI{400}{\kilo\metre}$ in the corona. In this sequence, \cite{Berghmans2021} automatically detected and cataloged 1467 brightening events, nicknamed "campfires" and referred to as "events" from hereon. The detection was performed after remapping the images on a regular $2400\times 2400$ Carrington grid spanning from \ang{248.9} to \ang{287.9} in longitude and from \ang{-11.5} to \ang{27.5} in latitude (corresponding to a \ang{0.01625} pitch, \SI{198}{\kilo\metre} on the sphere) and with a projection radius of 1.004~\(\textup{R}_\odot\) (Fig.~\ref{fig::field_of_view_label_modified}). As the spacecraft jitter had been documented in the FITS headers, it was compensated for in the Carrington remapping, and the absolute pointing values were determined by cross-correlation with AIA.

The automated detection scheme \citep[appendix B of][]{Berghmans2021} defined the events as pixels whose intensity is larger than an arbitrarily defined threshold of five times the local noise level in the first two smaller scales of a spatial à trous wavelet transform. Events overlapping between successive frames were merged to produce the final set of spatio-temporal events. Their surfaces range from \SI{0.04}{\mega\metre\squared} (the \hrieuv spatial resolution) to \SI{5}{\mega\metre\squared}, the upper limit being partly a consequence of the chosen maximum wavelet scale. No restriction was imposed on their duration. We note that the number of detected events, as well as their properties (surface, lifetime), depended highly upon the detection parameters. For consistency, we used the \citep{Berghmans2021} cataloged as is. We however removed the events present in the first or last image of the \hrieuv observation, as their lifetime might not have been fully captured. Fig.~\ref{fig::field_of_view_label_modified} (b) shows the location of the 1314 selected events on the first \hrieuv image of the sequence.

\subsection{Multichannel observations with AIA}

\begin{figure}[ht]
\includegraphics[width=9cm]{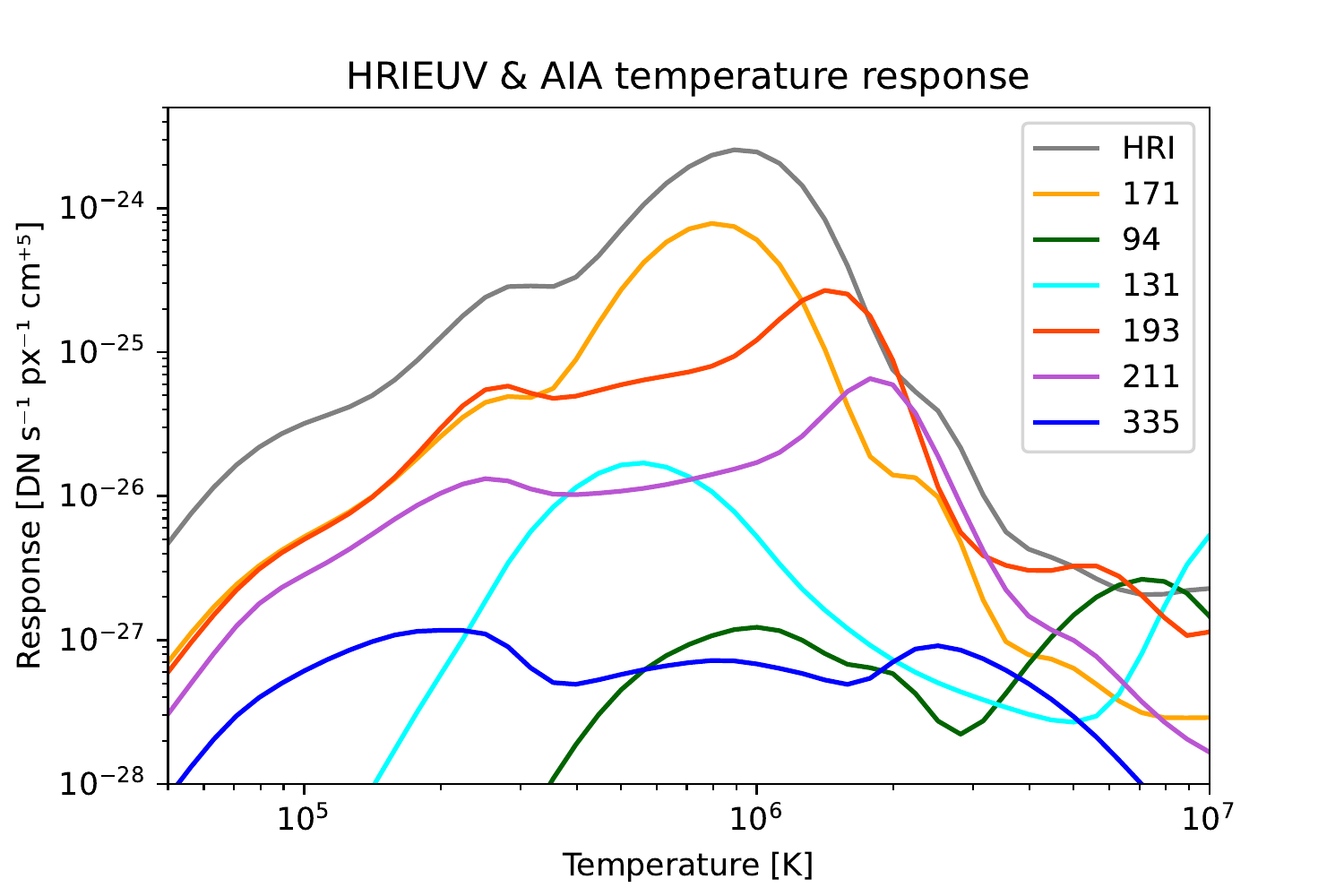}
\caption{\hrieuv and AIA temperature response functions computed with CHIANTI 10.0.1 \citep{1997A&AS..125..149D,DelZanna2021}, assuming an electron number density $n_e=\SI{e9}{\per\cubic\centi\meter}$.}
\label{fig::AIA_Tresponse}
\end{figure}

A major limitation of \hrieuv is its single passband, which makes it impossible to derive information on the plasma temperature. Therefore, we used data from six channels (94, 131, 171, 193, 211, and 335\,Å) of the AIA instrument for this purpose. We did not include the 304 band because the \ion{He}{II} \SI{30.4}{\nano\metre} spectral line is optically  thick, and the interpretation of its intensity is not straightforward. The selected bands cover a wide range of plasma temperatures (\SIrange{0.2}{8}{\mega\kelvin}, Fig.~\ref{fig::AIA_Tresponse}) but have only less than half the temporal resolution (\SI{12}\second) of \hrieuv (\SI5\second).

For our work, we needed to take into account the lower spatial and temporal resolutions of AIA, compared to those of \hrieuv. This difference meant that small and short-lived events detected by \hrieuv may be unresolved when observed with AIA 171. In addition, events might not be sufficiently bright in some of the AIA bands to be detectable.
The \hrieuv and AIA images were paired taking into account the \SI{229}\second\ difference in light travel time to Solar Orbiter and to the Earth. The AIA images were remapped onto the same Carrington grid as the \hrieuv data (Fig.~\ref{fig::field_of_view_label_modified}, d). On this common grid, the \hrieuv images were resampled with at least one grid point per pixel, and the AIA images were resampled with at least two.

\section{Method}
\label{method}

In order to characterize the evolution of the thermal structure of these events, we used the time lag method. Because the AIA bands peak at different temperatures (Fig.~\ref{fig::AIA_Tresponse}), the time lags between them are a signature of plasma cooling (or heating) over time. For example, the response functions of the AIA 193 and 171 bands peak respectively at \SI{1.5}{\mega\kelvin} and \SI{0.9}{\mega\kelvin}. The intensity in the 171 band peaking after  the 193 band can be interpreted as a hot plasma cooling, while the opposite behavior can be a signature of plasma heating. We discuss the various possible scenarios in detail in Sect.~\ref{discussion}.

In the following, we describe the computation and classification of the AIA light curves (Sect.~\ref{sec::LC}) and the computation of the time lags (Sect. \ref{sec::time_lag}). The analysis was performed pixel by pixel to take into account the spatial and the temporal information contained in the data. Several events were spatially resolved in the AIA data so that the thermal behavior in individual pixels of each event would be independently characterized. This avoids the assumption that the event has no thermal substructure. 
This method could involve the use of low SNR for some of the pixels, but this can be avoided by performing the analysis over the integrated intensity from the whole spatial extension of the event. However, the latter approach would impose the above-mentioned assumption, which we preferred to avoid. We verified in Appendix~\ref{ann::event_ana} that a same time lag analysis, performed over whole events, yields the same results.

Also in the following, we use "background" to refer to the total of background and foreground emission superimposed on the events along the line of sight. Background emission can represent a large fraction of the total emission (Sect.~\ref{sec::inpaint_not_inpaint}) and has the same properties as the QS emission observed outside the events. Since we wanted to measure the time lags of the events themselves, it was necessary to check the influence of the background (described in Section \ref{sec::LC}). The background intensity was estimated for each pixel and time step.

\subsection{Light curves}
\label{sec::LC}

\begin{figure*}[ht]
    \centering
    \includegraphics[width=\textwidth, trim=20 0 20 0, clip]{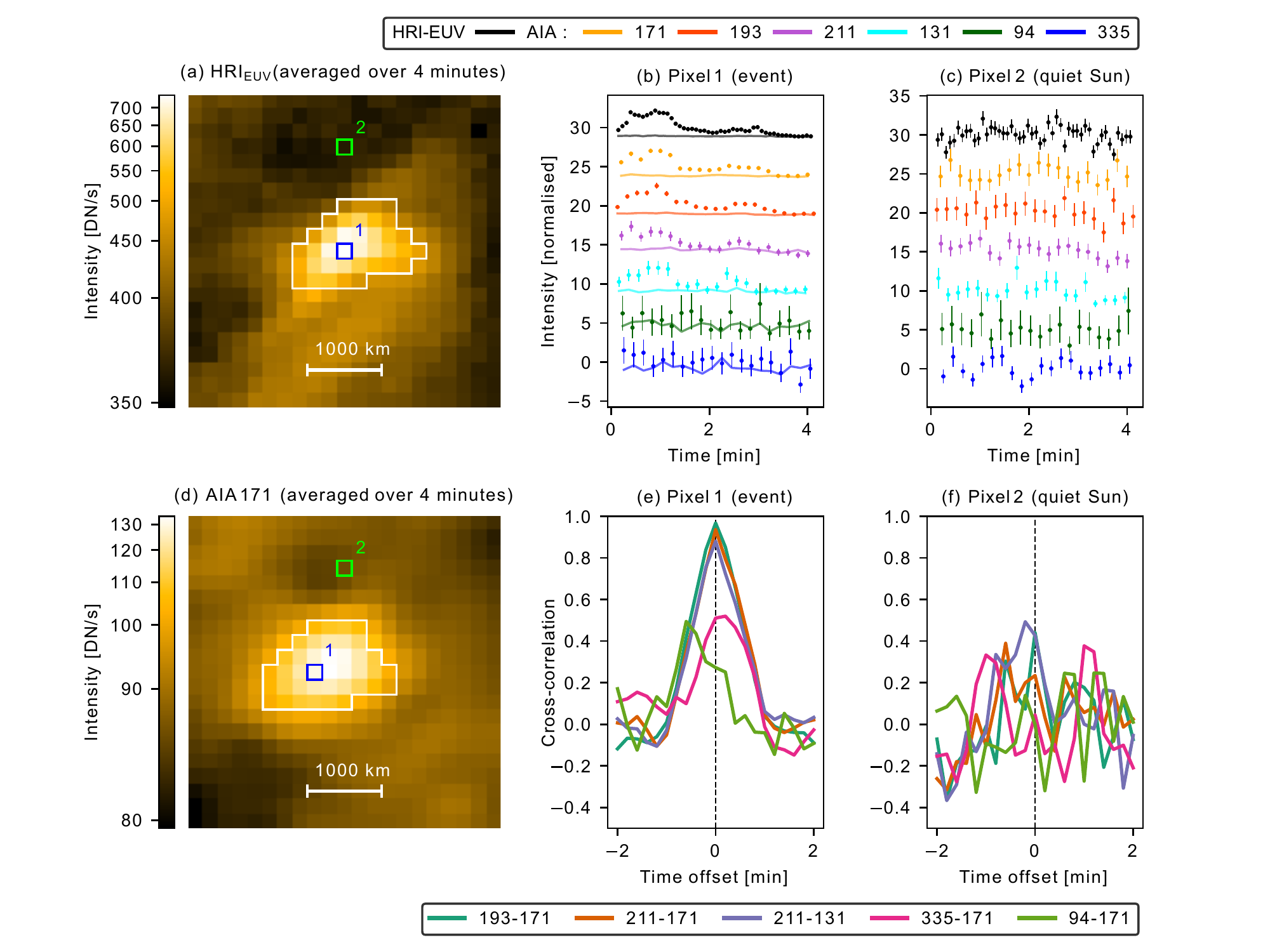}
    \caption{Illustration of time lag extraction method between AIA channels on event and QS pixels. Images of \hrieuv (a) (14:54:00 to 14:58:05 UT) and AIA 171 $\mathrm{\mathring{A}}$ (d) (14:57:45 to 15:01:57 UT) averaged in time over their respective sequence on May 30, 2020. Both images are centered around Carrington coordinates (275.00, 9.07)°. The white contours represent the masks that isolate the event pixels from the QS ones.  Pixels 1 and 2 were selected, respectively, as an example for the event pixel and QS pixel. Subfigure (b) shows the light curves in pixel 1 for \hrieuv and the AIA channels original data (dots) and background data estimated with the "inpainting" algorithm (solid curves). For each channel, both curves are normalized over the standard deviation over time of the original data (dots). 
    Subfigure (c) shows the light curves in pixel 2 for the same channels of (b) normalized to their standard deviation over time.
    Different couples were separated by an arbitrary value of five. 
    The error bars in subfigures (b) and (c) were computed from the shot and read noises.
    (e) and (f) show the cross-correlation as a function of the time offset between the AIA light curves for pixels 1 (b) and 2 (c), respectively.}
    
    \label{fig::event3_test_v2}
\end{figure*}

For our analysis, we classified the pixels into two categories: "event" pixels, that is, those containing at least one event from \citet{Berghmans2021} during the sequence, and non-event pixels that we refer to as "QS" for simplicity. The QS pixels were used as a reference, and their statistics were compared to those of the event pixels (Sect. \ref{sec::inpaint_not_inpaint}).

While the AIA and \hrieuv data were reprojected to the same Carrington grid, the location of each event could be different in the two data sets. Indeed, the separation angle between the two vantage points induced a parallax shift for those events located above or below the projection sphere. The contour of each event detected in \hrieuv was shifted by the amount measured by \cite{Zhukov2021} to obtain the corresponding contour in AIA. In the case of spatially overlapping events, this can cause the classification mask (the union of the contours at each time step) in AIA to have a different shape  than in \hrieuv. This is the case for the area shown in Fig.~\ref{fig::event3_test_v2}, in which two successive events peaking at 14:54:30 and 14:55:04 UT are overlapping and do not have the same height and thus do not have the same parallax shift.

We estimated the background emission at each pixel using the \verb|open-cv| implementation of the inpainting method of \cite{990497}. This method estimates the intensity inside the mask by matching the intensity and intensity gradients at the boundary of the mask. This operation was performed at each time step. Whenever this background subtraction was applied to the analysis, we mentioned so explicitly in the text.

Figure \ref{fig::event3_test_v2} (b) shows an example of the result from this treatment. We have selected a pixel inside the mask (pixel 1 in Fig.~\ref{fig::event3_test_v2} (a)), and we plotted the light curves as measured in the \hrieuv and AIA channels together with their calculated  background emission. 
For display purposes, original and background-subtracted light curves were normalized to the standard deviation of the original.
To plot all the curves on the same panel, we separated the curves from a given channel vertically by an arbitrary value of five.
The error bars are the root sum square of the photon shot noise (as computed in Appendix~\ref{sec::confidence}) and read noise components. \cite{Boerner2012} provides the read noise for all the AIA bands. For \hrieuv, the read noise is estimated to be equal to 1.5~DN.
In Fig.~\ref{fig::event3_test_v2} (b), the light curves of all channels but AIA 94 and 335 have a similar behavior. In the AIA 94 and 335 channels, the event in pixel 1 was not detected above the noise. The absence of signal in these two bands is caused by their low response (see Fig.~\ref{fig::AIA_Tresponse}) and is common for most of the events.

Figure \ref{fig::event3_test_v2} (c) shows, for a comparison, the same as Fig.~\ref{fig::event3_test_v2} (b) but for a representative QS pixel (pixel 2 in Fig.~\ref{fig::event3_test_v2} (a)). Apparently uncorrelated fluctuations of the intensity can be seen in this figure.

\subsection{Time lags}
\label{sec::time_lag}

\begin{figure*}[ht]
    \hspace{-1.2cm}
    \includegraphics{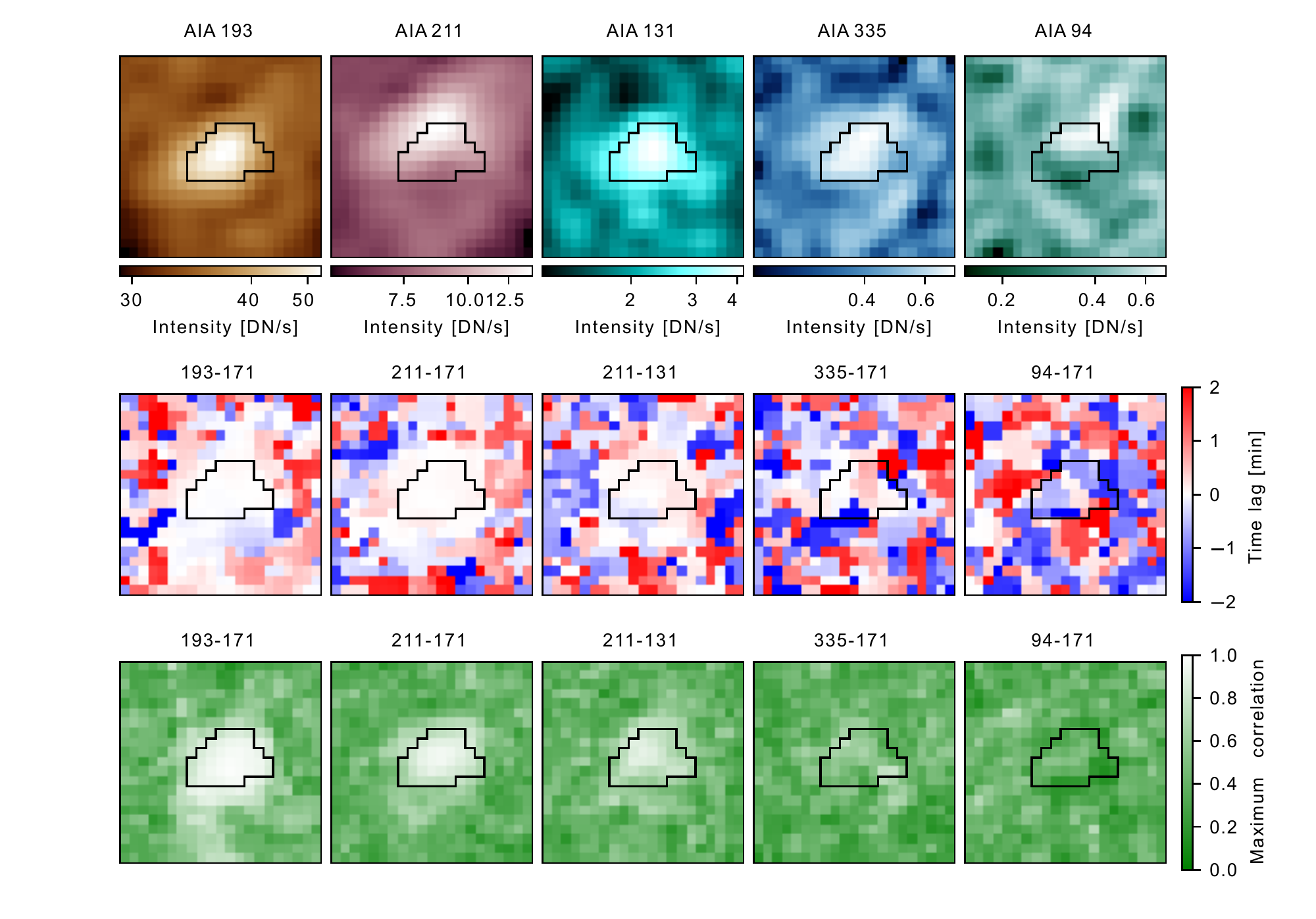}
    \caption{Time extraction procedure applied pixel by pixel to event pixels and their surrounding QS pixels. Top row: Intensity maps for five AIA  bands (averaged over the temporal sequence) showing the event of  Fig.~\ref{fig::event3_test_v2}~(a) and (d). The "event" region is identified by the black contour. Middle and bottom rows: Time lag and associated maximum correlation maps for five couples of AIA bands. These are the result of the pixel-by-pixel cross-correlation analysis. 
    The maximum correlations of the events decreases as the intensities of the involved AIA channels decrease.}
    \label{fig::time_lag_max_corr_maps}
\end{figure*}

In the following subsection, we describe the computation of time lags between couples of AIA light curves. The time lags are defined as the temporal offset between the two light curves that yields the maximum Pearson's cross-correlation coefficient.

By design, the images of the six channels are not cotemporal. For this reason, we resampled the light curves on the timeline of the 171 band using linear interpolation before applying the cross-correlation procedure. The latter was performed on a range of temporal offsets of $\pm2$ minutes with steps of 12\,s. A finer estimate of the time lag was obtained by parabolic interpolation around the maximum.

Figures~\ref{fig::event3_test_v2} (e) and (f) show the results of this analysis for the AIA pixels 1 and 2. We plotted the values of the correlation as a function of the time offset applied between the two light curves. For the event pixel, we chose  three couples with high SNRs (193 -- 171, 211 -- 171, and 211 -- 131). These couples have a strong correlation peak at near-zero offsets: \SI{0.3}{\second} for 193 -- 171, \SI{0.8}{\second} for 211 -- 171, and \SI{-0.6}{\second} for 211 -- 131.

The other two curves (335 -- 171 and 94 -- 171) involve low SNR bands and have a maximum of correlation at a time offset different from zero. Their time offset is positive for 335 -- 171 (\SI{7.8}s) and negative for 94 -- 171 (\SI{-3.2}s), with a maximum correlation below 0.5. The SNR is low in the 335 and 94 bands, and the peak correlation is of the order of that found in the QS (Fig.~\ref{fig::event3_test_v2} (f)). We discuss the significance the cross-correlations involving low SNR bands in Sect.~\ref{sec::short_time_scale_coolings} and~\ref{sec::max_corr_signal}.
Figure \ref{fig::event3_test_v2}~(f) shows the results for the selected QS pixel. Clearly there is no strong correlation at any time offset and for any pair of AIA channels.

Figure \ref{fig::time_lag_max_corr_maps} displays the maps of AIA intensity averaged over the sequence, time lag,  and maximum of cross-correlation for the area shown in Fig.~\ref{fig::event3_test_v2}. We noticed that the emission is not cospatial in all bands. The intensity maps  show a displacement of emission peak for AIA 211 and 94 (even though the signal is very low for AIA 94). Since the AIA channels are all coaligned, this could be due to the thermal structure of the observed features. 
These observations show the importance of analyzing the plasma evolution pixel by pixel, as opposed to averaging the intensity over the event surface. While doing the latter might increase the SNR, it would mix light curves of regions at different temperatures. 

The bands in the top row of Fig. \ref{fig::time_lag_max_corr_maps} are ordered by decreasing mean intensity and thus decreasing SNR. In the bottom row, within the mask, we see correspondingly decreasing correlation values. Higher correlation values are associated with spatially coherent near-zero time lags, whereas lower correlations show an apparently random distribution of the time lags. 

\section{Results}
\label{results}

In Sect. \ref{sec::short_time_scale_coolings} we present the statistical analysis over the whole field of view of the data. In Sect. \ref{sec::max_corr_signal}, we discusses the effect of the SNR on the results. Finally, we  estimate the effect of the background on the time-lag analysis in Sect. \ref{sec::inpaint_not_inpaint}.

\subsection{Zero time lags}
\label{sec::short_time_scale_coolings}

\begin{figure*}
    \centering
    \includegraphics[width=\textwidth]{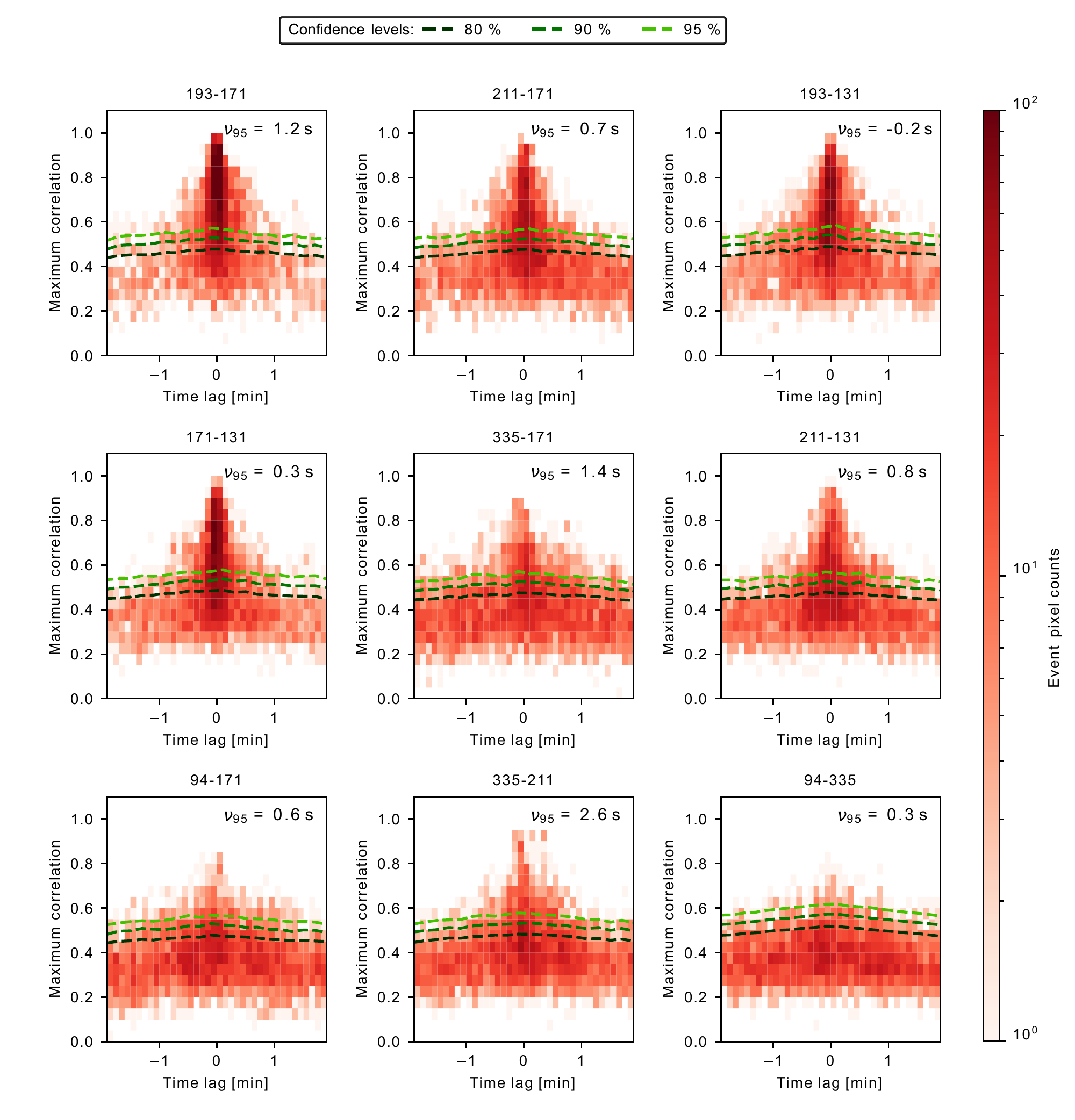}
    \caption{Two-dimensional histograms (shades of red) of time lags and maximum correlations of nine couples of AIA channels for the 4451 event pixels of the \hrieuv field of view.
    The estimated background has been subtracted. The green dashed lines are the confidence levels, derived in Appendix~\ref{sec::confidence}. The $\nu_{95}$ parameter quantifies the asymmetry of the time lag distributions. It represents the average of the event time lags above the 95\% confidence level weighted by their respective maximum correlations. } 
    \label{fig::test_figure_time_lags}
\end{figure*}

For the event pixels, Fig.~\ref{fig::test_figure_time_lags} displays the time lag and the maximum correlation 2D histograms. We chose nine representative AIA couples covering a wide range of temperature sensitivities. In this part of the work, the estimated background has been subtracted from the event pixel intensity.

The 80\%, 90\%, and 95\% confidence levels, displayed in Fig.~\ref{fig::test_figure_time_lags}, are computed in Appendix \ref{sec::confidence}. 
The counts above the 95\% level are at most 5\% likely to occur by chance. For most of the couples, a significant number of pixels are centered around short time lags (below the 12 s cadence), and are above the \SI{95}{\percent} confidence level in cross-correlation. This part of the distribution is therefore statistically significant.
In contrast, 94 -- 335 shows no significant pixel counts above the \SI{95}{\percent} confidence level, which matches the contour of the 2D histogram. Given that these bands are largely affected by noise, this validates, a posteriori, the principle of computing confidence levels from uncorrelated light curves (Appendix \ref{sec::confidence}).

While the time lags are near zero, the distributions are slightly asymmetric. This can be quantified by the parameter $\nu_{95}$, which represents the average of the time lag values above the 95\% confidence level weighted by their maximum correlation. Apart from the 335 -- 211 couple, all the asymmetries are below the exposure time of \SI{2}{\second}. For 335 -- 211, the positive asymmetry is above the exposure time but below the temporal resolution.

\subsection{Influence of the signal level}
\label{sec::max_corr_signal}
\begin{figure*}
    \centering
    \includegraphics[width=\textwidth]{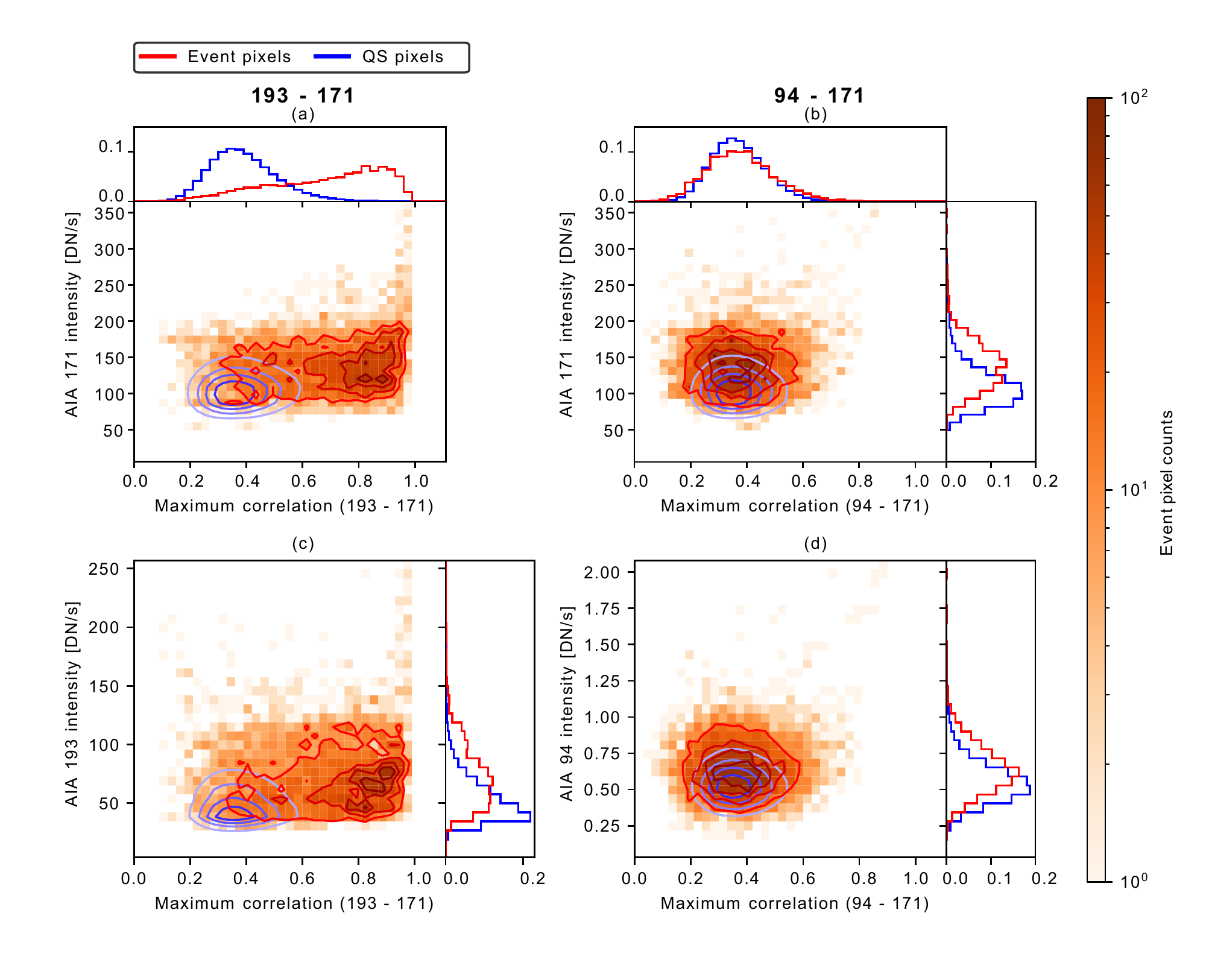}
    \caption{ Main panels: Histograms of the time-averaged intensity, as a function of the maximum cross-correlation, in the whole \hrieuv\ field of view. The left and right columns show the results for the 193 -- 171 and the 94 -- 171 couples, respectively.
    The 2D orange histograms are the counts of event pixels. The 2D red and blue contours correspond to the 20, 40, 60, and 80  percentiles of the events and the QS pixel distributions, respectively. The histograms in the margins were normalized by their total number of counts. No histogram is displayed in the right margin of (a), as it would be a repetition of that of (b). Similarly, the top-margin histograms of (c) and (d) have been omitted, as they are the same as those of (a) and (b), respectively.}
    \label{fig::max_corr_vs_signal_multiple}
\end{figure*}

The main panels of Fig.~\ref{fig::max_corr_vs_signal_multiple} display the 2D histograms of the average intensity over the time sequence versus the maximum correlations for the two AIA couples 193 -- 171 (high-high SNR) and 94 -- 171 (low-high SNR). The bottom and top rows, respectively, display the intensity of the first and second band of the pair. Both distributions of event and QS pixels are displayed in Fig.~\ref{fig::max_corr_vs_signal_multiple}.
The orange distributions and red contours refer to the event pixels, and the blue contours refer to the QS pixels. For the 193 -- 171 couple (Fig. \ref{fig::max_corr_vs_signal_multiple} (a) and (c)), the event pixel distribution shows a wide range of possible correlation values, in contrast to the QS pixel distribution. The latter is more compact and centered around lower maximum correlation and intensity values. However, the 171 -- 94 case shows both the event and the QS pixel histograms as sharing a similar compact shape. This is mostly due to the lower intensity and thus lower SNR in the 94 band.

The intensity distributions, which are displayed in the histograms in right margins  of Fig.~\ref{fig::max_corr_vs_signal_multiple}, peak at higher values for the event pixels than for the QS pixels for every channel. This implies that, on average, the \hrieuv events are also visible in the AIA channels.
The most significant difference between the two AIA couples shown in the figure is their maximum correlation distributions, displayed in the top-margin histograms. Indeed, while the event pixel distribution peaks at higher correlation values than the QS distribution for 193 -- 171, both distributions share a similar shape for 94 -- 171. 

As shown in the intensity distributions of the right-margin histograms, the signal in the  94 band is much lower than in the other two bands. Given the exposure time of 3 s, the 94 band intensity distributions (Fig.~\ref{fig::max_corr_vs_signal_multiple} (d)) are close to the read noise value of 1.14\,DN. The SNR of the median intensity over the QS in the field of view is 13.7, 9.5, and 0.7 for the 171, 193, and 94 bands, respectively. Thus in the 94 band, the noise dominates, and the events, if present in the band, remain undetected for most of the cases (see Fig. \ref{fig::event3_test_v2} (b) as an example). 
This is why in the 94 -- 171 case the maximum correlation distributions of the events and the QS pixels share the same statistical behavior, as most of the signal in this band originates from the noise. Figure \ref{fig::test_figure_time_lags} mostly shows non-significant time lags resulting from noise for the couples 94 -- 171 and 94 -- 335.

\subsection{Influence of the background subtraction}
\label{sec::inpaint_not_inpaint}
\begin{figure*}[ht]
    \hspace{-1cm}
    \includegraphics[width=1.04\textwidth]{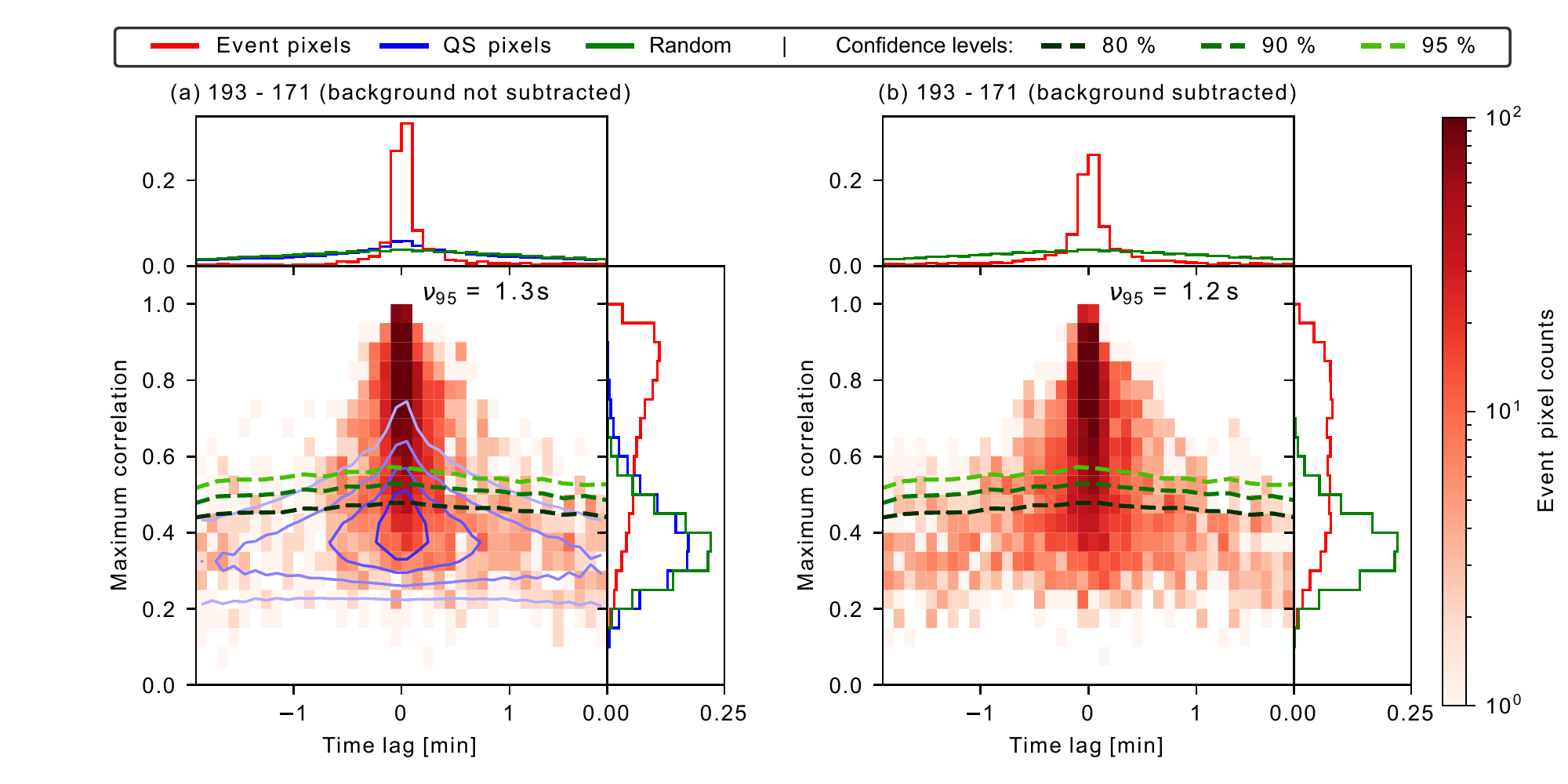}
    \caption{Margin and 2D histograms of time lags and associated maximum correlation values for couple 193 --171. Subfigure (a) in red shows the original distribution for the event pixels. Subfigure (b) shows  the background subtracted event pixels.  
    The blue contours in the central panel of subfigure (a) are the 20, 40, 60, and 80 percentiles of the QS pixel distribution. 
    The green colors in the main panels are the confidence levels, and the distributions of the light curves used to compute them are plotted with the same color in the margin histograms. The margin histograms were normalized by their total number of pixels. The parameter $\nu_{>95}$ is defined as in Fig.~\ref{fig::test_figure_time_lags}.}
    \label{fig::time_lag_vs_max_corr}

\end{figure*}

According to Fig. \ref{fig::max_corr_vs_signal_multiple}, the maximum of the AIA 171 intensity distribution is only about 1.3 times higher for the event pixels compared with the QS pixels. Therefore, the background largely contributes to the overall signal. This is why it is necessary to evaluate the influence of the background on the cross-correlations, which we illustrate using the couple 193 --171. Figure~\ref{fig::time_lag_vs_max_corr} displays the time lag and the maximum correlation distributions without and with the subtraction of the estimated background component (as described in Sect.~\ref{sec::LC}). 
Both the event and the QS pixel distributions are included. In the margins, the 1D event pixels distributions are displayed in red and the QS ones in blue. The green histograms correspond to the uncorrelated light curves used to compute the confidence levels (Appendix \ref{sec::confidence}).

As in Fig.~\ref{fig::test_figure_time_lags}, the event distributions peak at high correlation values and are concentrated around short time lags. The impact of the background intensity on the event distributions is visible when comparing Fig.~\ref{fig::time_lag_vs_max_corr} (a) with (b); the time lags and their asymmetries are mostly unchanged.

However, when removing the background, the distribution of the event pixels becomes flattened (most visible in the margin histogram), and the counts are redistributed in the low-correlation random time lag wings. This outcome has two causes. First, the noise from the QS is propagated to the background by the inpainting (Sect.~\ref{sec::LC}) and in turn to the background-subtracted light curves. Thus, the correlations are lowered in this case. Second, the QS signal is partly correlated around the zero time lag. This forms the high-correlation tail visible in the blue contours of Fig.~\ref{fig::time_lag_vs_max_corr}~(a). Subtracting the background removes this correlated signal and also lowers the correlations.
To conclude, removing the background isolates the contribution of the events to the time lags. Thus, the time lags in Figures~\ref{fig::time_lag_vs_max_corr}~(b) and~\ref{fig::test_figure_time_lags} are a property of the event pixels and not of the QS.

\section{Discussion}
\label{discussion}
In this work, we have presented the results from the statistical analysis of the time lags measured in the AIA data for the small-scale EUV brightening events (the so-called campfires) cataloged by \cite{Berghmans2021}. This catalog has the unique property of collecting the tiniest and most rapid brightening ever observed, which are the manifestation of  physical processes probably already known  but now observed over shorter temporal and spatial scales. For this reason, we preferred to use the general name of EUV events.

Our observational work points to the following result: The EUV events are characterized by short time lags (within $\pm 12$ s) and high correlations. We verified that these results are statistically significant and are not caused by background variations alone. In comparison, the QS mostly exhibits random time lags with lower correlations. It is possible that the timescales of thermal changes between events and the surrounding areas are different, the latter being much longer than the maximum time lags considered here.

To our knowledge, this is the first time that the time lags associated with small-scale EUV brightenings and their surroundings have been statistically characterized. Earlier works, as mentioned in the introduction, that used the time lag technique reported zero time lags in the QS surrounding active region loops without taking into account the possible presence of small-scale brightenings.\\

Concerning the interpretation of the short time lags, there are three possible scenarios that can be raised: (1) the observed events do not reach the peak temperatures of the response function ($\sim$ one million degrees); (2) The observed events reach coronal temperatures ($> 1$ MK), but their fast cooling, subpixel multithermal structure, and the width of the AIA response function prevent us from detecting significant time lags; and (3) the observed events are the transition region ($\sim$ 1MK) emission of long and hot \citep[i.e., $\sim$ 10 -- 100 Mm, $\sim$ 3 MK;][]{Reale2014} loops, which are heated impulsively.

Starting with the interpretation given by the first scenario, looking at the most intense bands of AIA (Figure \ref{fig::AIA_Tresponse}), we understand that 
a time lag of zero arises when the plasma temperature does not reach the peak of the 171 band. At the temperatures below this peak, all the bands behave similarly and so do the light curves. Furthermore, the observational properties (low-lying, short time lags) of these events resemble what is observed by \cite{Winebarger2013} for the Hi-C loops ($T_e \sim 10^5$ K and $n_e \sim 10^{10} \ \mathrm{cm^{-3}}$) in the inter-moss loop areas.
Their time lag analysis on the AIA light curves also displayed near-zero time lags, which led them to conclude that the loops did not reach one million degrees. They interpreted their observations as loops being heated impulsively with low-energy nanoflares. These loops would then cool rapidly because of their short length. Given the similarities of these Hi-C loops with the \hrieuv events,  we suggest that they may have a similar physical origin, that is, being the result of impulsive heating.

For cold events to be visible in the  AIA bands and in \hrieuv, they should be quite dense.  
We did a first order estimation of density of the \hrieuv events using an average value of the background-subtracted event intensity on AIA 171 and assuming an isothermal plasma. We obtained $n_e \sim \SI{e9}{\per\centi\meter\cubed}$ for $T_e = \SI{1.3e6}{K}$ and $n_e \sim \SI{e10}{\per\centi\meter\cubed}$ for $T_e = \SI{3e5}{K}$. The latter supports the result of \cite{Winebarger2013}. 

However, we had to consider possible differences between the Hi-C loops and our \hrieuv events. First, as mentioned, the observed solar region is not the same. But small low-lying cool loops ($T_e \le$ 0.5 MK) are observed in the QS \citep{Hansteen2014} and are ubiquitous along the supergranular cell boundaries in the upper solar atmosphere \citep[see for instance,][and references therein]{feldman1999, Almeida2007}. And since there is no distinction between supergranular cells in QS and active regions, we expected to observe similar events in both regions. \cite{Berghmans2021} showed with \hrilya observations that the \hrieuv events are organized mostly around the supergranular network.

Another difference between the Hi-C and the \hrieuv events are their estimated temperature, around $T_e \approx 0.25 \pm 0.06$ MK for Hi-C events \citep{Winebarger2013} and around $1.3 \pm 0.1$ MK for \hrieuv events \citep{Berghmans2021}. In case these are similar events, we suggest that the above  difference  in temperature may be the result of the uncertainties in the data, on the adopted inversion methods (which is different for the two analysis), and the associated assumptions which are applied to relatively broad band instruments, as for these two imagers. Indeed, measuring the temperature of these events is very challenging.
For instance,  \cite{Schonfeld_2020} showed that the cool plasma emission often dominates the bands even though the hot plasma is present.

Assuming the second scenario, a time lag close to zero for AIA bands has been predicted by \citet{Viall_2015} in the TR emission of active region coronal loops heated by nanoflares. These authors
showed that the combination of the multi-temperature sensitivity of the AIA bands combined with the almost constant pressure property of the TR and its variable extension along the loop during the heating-cooling phases result in a narrower time lag with respect to the coronal emission part of the loop. We emphasize here that the TR of a loop is defined as the region where the thermal conduction acts as a plasma coolant, contrary to the coronal region where it acts as a heater \citep[e.g.][]{Klimchuk2008}.
While the presence of short time lags for all the AIA couples in the simulation corroborates with our results, the loops modeled by \cite{Viall_2015} are much longer than what we observed ($L \approx$ 30 -- 50 Mm, with respect to 0.4 -- 4 Mm).
Moreover, in those simulations, a clearly different signature in the time lag exists between TR and coronal emission, while this characteristic is not visible in our data. This could possibly be explained by the short cooling time from coronal temperatures of one of these tiny loops. For instance, for the shorter loops ($\sim$ 0.4 Mm) detected by \hrieuv at a temperature of $\sim$ 1.3 MK and density of $n_e=10^{10} \mathrm{cm^{-3}}$, the cooling time is about 14 s.

Due to the AIA cadence of 12 s, it is possible that our time lag method is not sensitive enough to detect both TR and coronal emission populations of short time lags. We propose to further investigate this aspect in the future through numerical simulations. 
The small asymmetries we have in our time lag distributions are below the cadence of our observation, and we would need data with a higher temporal resolution to corroborate such a result. The cadence should be at least similar to the one of \hrieuv, where the emission variation of the event is better captured. At present, we verified that the measured time lags are independent of the event's duration.

Concerning the third scenario, if such large loops exist in the QS, they remain undetected by the AIA channels, meaning that they would have a very low density. Without independent evidence that this is the case, we exclude this possibility for now.

In conclusion, in the picture of impulsive heating phenomena acting in the QS region and considering the wide temperature response of the AIA bands, our results appear to also be consistent with predominantly fast cooling plasma from more than 1 MK, satisfying our scenario two interpretation. 
Consistent with this picture are the results from a 3D magneto-hydro-dynamics (MHD) simulation using Max Planck für Sonnensystemforschung/University of Chicago Radiative MHD (MURaM) code by \cite{chen2021}. 
In those simulations, magnetic reconnections in the coronal part of small QS loops produced events with properties similar to what was observed in \hrieuv.
The authors noticed that the simulated \hrieuv emission only showed the apex of the heated loop, where the lower density allows the available stored energy to heat the plasma up to $\approx 1.3$ MK, even though some hotter temperatures could also be reached.

To summarize, our results are consistent with two possible scenarios: Either the events do not reach coronal temperatures or they do but they cool faster than the AIA temporal resolution. It is possible that the two scenarios coexist, as the \hrieuv catalog does not separate events produced by different physical processes.
The AIA cadence and the multithermal nature of the bands do not allow separating the emissions from the possible cool and hot plasma along the line of sight.

To solve the ambiguity in the temperature, we need to use spectroscopic data. 
This has been done recently by using the {Spectral Imaging of the Coronal Environement (SPICE)} instrument on board Solar Orbiter \citep[][]{Huang2022}. 
The authors investigated a few HRIEUV events and came to the conclusion that the studied events do not show significant emission at temperatures higher than that of Ne~{\sc viii} (0.63 MK).

Although such spectroscopic analysis needs to be extended to a larger sample to better quantify the fraction of events not reaching high temperatures, we find that this analysis supports our conclusion that QS small-scale EUI brightenings are in most cases largely dominated by  cool emission. Further investigations are needed to confirm this idea. For these reasons, we plan to extend our methodology to forward modeling constrained by spectroscopic data.

\begin{acknowledgements}
The authors gratefully thank J.A. Klimchuk for the fruitful discussions and suggestions. 
The authors thank the referee for the constructive comments that helped to improve the manuscript.
A.D. acknowledges the funding by CNES and EDOM.
S.P. acknowledges the funding by CNES through the MEDOC data and operations center. 
G.P. was supported by a CNES postdoctoral allocation.
P.A. and D.M.L. are grateful to the Science Technology and Facilities Council for the award of Ernest Rutherford Fellowships  (ST/R004285/2 and ST/R003246/1, respectively). The ROB team thanks the Belgian Federal Science Policy Office (BELSPO) for the provision of financial support in the framework of the PRODEX Programme of the European Space Agency (ESA) under contract numbers 4000134474 and 4000136424.
This paper uses the Solar Orbiter/EUI data release 1.0 \url{https://doi.org/10.24414/WVJ6-NM32}.
Solar Orbiter is a space mission of international collaboration between ESA and NASA, operated by ESA. The EUI instrument was built by CSL, IAS, MPS, MSSL/UCL, PMOD/WRC, ROB, LCF/IO with funding from the Belgian Federal Science Policy Office (BELSPO/PRODEX PEA 4000134088, 4000112292, 4000117262, and 400013447); the Centre National d’Etudes Spatiales (CNES); the UK Space Agency (UKSA); the Bundesministerium für Wirtschaft und Energie (BMWi) through the Deutsches Zentrum für Luft- und Raumfahrt (DLR); and the Swiss Space Office (SSO).
This work used data provided by the MEDOC data and operations centre (CNES / CNRS / Univ. Paris-Saclay), http://medoc.ias.u-psud.fr/. This research used version 0.6.4 \citep{barnes5606094} of the \verb+aiapy+ open source software package \citep{Barnes2020}. \end{acknowledgements}

\bibliographystyle{aa}
\bibliography{biblio.bib}

\begin{appendix}

\section{Computation of the confidence levels}
  \label{sec::confidence}

The cross-correlation of two uncorrelated, random time series has a nonzero probability of resulting in a time lag with a nonzero value for the maximum correlation. 
This is why the interpretation of our time lag results is challenging, especially for couples involving low- to medium-SNR AIA channels, such as 131, 94, and 335, which are noise dominated in several pixels. 

For our purpose, we adopted a Monte-Carlo approach inspired by \cite{10.1093/mnras/stu1707}. We computed the time lags (corresponding to the maximum cross correlation) between many uncorrelated, simulated AIA light curves to estimate the probability of chance occurrence of each time lag value.

The simulated light curves were built using the observational results that  showed the coronal emission to have a temporal power spectral density (PSD) that can be modeled by a power law \citep{auchere_long-period_2014, gupta_observations_2014, 2017SoPh..292..165T}. Specifically for the QS pixels, \cite{Ireland_2014} fitted the exponents $n=1.72 \pm 0.01$ for AIA 171 and $n=2.20 \pm 0.01$ for AIA 193. To keep the empirical model simple, we adopted a power law with the exponent $n=2$ for all of the AIA channels. From this PSD, we generated $10^5$ random light curves of \SI4\minute\ in duration and \SI{12}\second\ cadence using the method described in \citep{Timmer&Koenig1995}. The obtained time series $\hat{I}$(t) (in arbitrary units) were converted into digital numbers (DN): 
\begin{equation}\label{as::hat_I}
    I(t)\ [\mathrm{DN}] = \left(\hat{I}(t) - \mu_{\hat{I}}\right)\frac{\sigma_\mathrm{DN}}{\sigma_{\hat{I}}} + \mu_\mathrm{DN}
\end{equation}
where $\mu_{\hat{I}}$ and $\sigma_{\hat{I}}$ are, respectively, the mean and standard deviation of $\hat{I}$(t); 
$\mu_\mathrm{DN}$ and $\sigma_\mathrm{DN}$ are the spatial mean intensity and the standard deviation derived from the first image of the AIA sequence (see Fig.~\ref{fig::field_of_view_label_modified} (d)).

 Photon noise was then added by picking random values from a Poisson distribution peaking at the average photons per image. We assumed this photon average to be equal to the incident photons I(t). Negative intensity values were set to zero.
Next, we simulated the regular AIA acquisition chain by reconverting the time series into DN.
Read noise was then added, in the form of a normal distribution of mean zero and standard deviation $\sigma_{RN}$. Using the inverse of the camera gain, $I(t)$ was converted into photons. All the conversion constants were taken from the initial AIA calibration \citep{Boerner2012}.
The resulting time series were then used for the time lag analysis applied to  each of the AIA couples in  Sect. \ref{sec::short_time_scale_coolings} and \ref{sec::inpaint_not_inpaint}. 
 
The time lags and maximum correlation distributions of these random light curves are displayed for the couple 193 --171 in the margin histograms of Fig.~\ref{fig::time_lag_vs_max_corr}. The confidence levels are defined as percentiles (80\%, 90\%, 95\%) of the maximum correlation distribution. According to our simulation, counts above the 95\% confidence level are at most 5\% likely to be caused by chance.

\section{Event-based time lag analysis}
  \label{ann::event_ana}
  \begin{figure*}
      \centering
      \includegraphics[width=\textwidth]{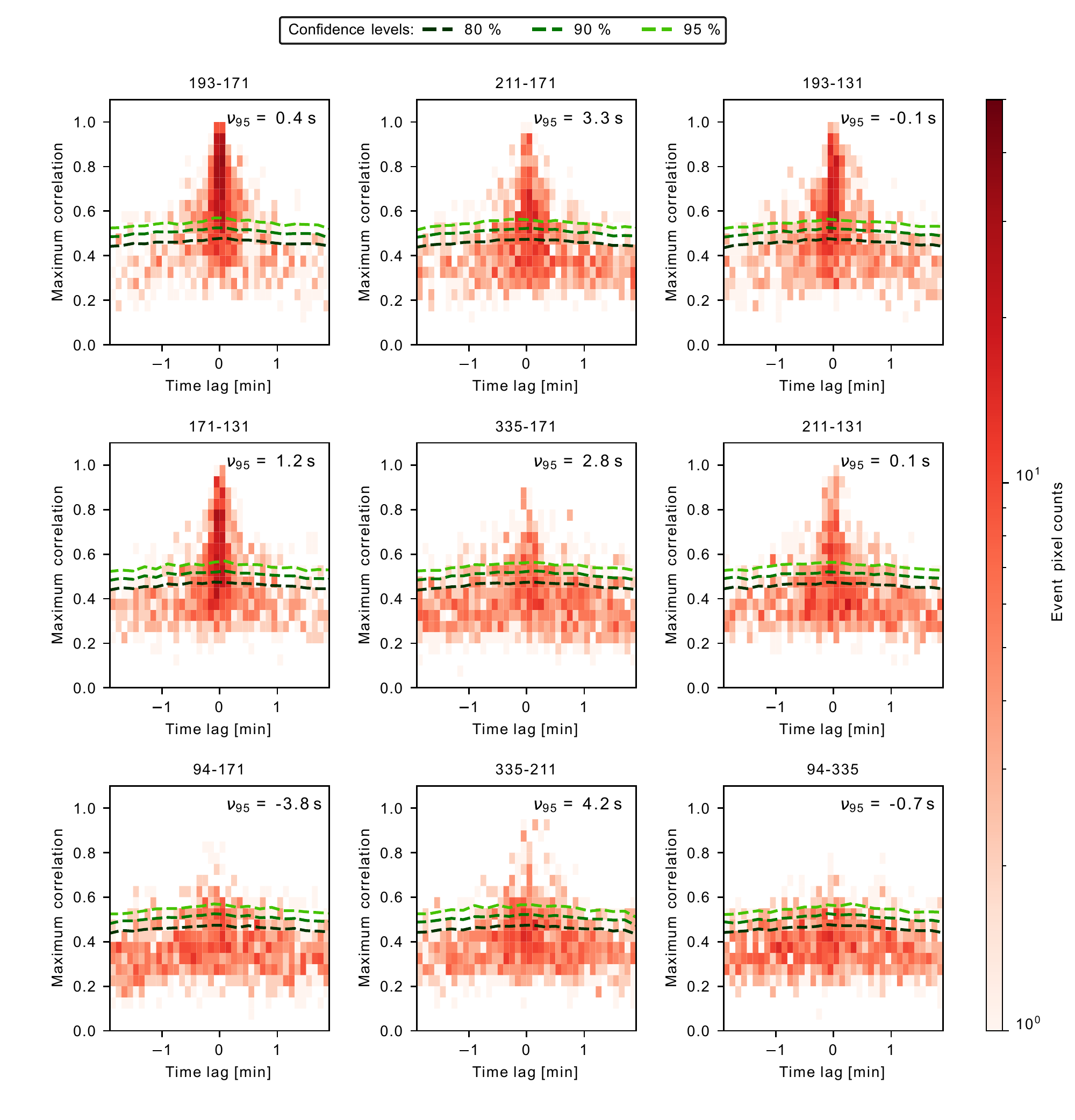}
      \caption{Same as Fig.~\ref{fig::test_figure_time_lags} but with a full-event approach, as opposed to a pixel-based one. For each of the 1314 events, the light curves were spatially averaged over each of their respective event surface. Then, time lag extraction was performed, similar to the pixel-based approach (Sect.~\ref{sec::time_lag}). The estimated background has been subtracted on event pixels with the "inpainting" algorithm.}
      \label{fig::multi_time_lags_full_event}
  \end{figure*}

The main work we have presented is based on the single-pixel analysis. Here we summarize the results from  the full-event investigation in order to verify if the resulting thermal behavior reflects the one deduced with the single-pixel analysis. 

The pixel-based and the full-event approaches both have their advantages. The full-event approach increases the SNR of the light curves, as it is represented by the averaged intensity over the selected event area,
but it does not separate the "cold" and the "hot" pixel populations. This is because inside an "event surface," one pixel might reach a higher temperature than the others. The high temperature pixel and the lower temperature pixel appear as separate counts in the resulting figures of the pixel-based approach (Fig.~\ref{fig::test_figure_time_lags}). In contrast, the temperature associated with the average intensity will be something  between the hotter and cooler pixels, reducing the temperature excursion over time. Each event area is a single count in the statistical analysis (Fig.~\ref{fig::multi_time_lags_full_event}).

To build the single-event light curves, we proceeded by spatially averaging the light curves within each event mask. The time lag analysis was then applied to these new time sequences in the same way as it was done for the pixel-based approach (Sect.~\ref{sec::time_lag}).

The results of the  analysis are displayed in Fig.~\ref{fig::multi_time_lags_full_event}. The time lags are centered around short values (>12s) above the 95\% confidence level. There is no noticeable difference with the pixel-based approach (Fig.~\ref{fig::test_figure_time_lags}), apart from small variations in the asymmetries $\nu_{95}$, which remain close to the exposure time. 
The variations are probably caused by the lower number of counts above the 95\% confidence level relative to the pixel-based approach.
Using the full-event approach decreases the statistical significance of the asymmetry, and the events should be studied individually.

\end{appendix}
\end{document}